# Phonons and Anisotropic Thermal Expansion Behaviour of NiX (X = S, Se, Te)

Prabhatasree Goel[1], M. K. Gupta[1], S. K. Mishra[1], Baltej Singh[1,2], R. Mittal[1,2], P.U. Sastry[1,2], A. Thamizhavel[3] and S. L. Chaplot[1,2]

[1]Solid State Physcis Division, Bhabha Atomic Research Centre, Trombay, Mumbai 400085, India
[2]Homi Bhabha National Institute, Anushaktinagar, Mumbai 400094, India
[3]Tata Institute of Fundamental Research, Homi Bhabha Road, Colaba, Mumbai 400005, India

Metal Chalcogenides have been known for important technological applications and have attracted continuous interest in their structure, electronic, thermal and transport properties. Here we present first principles calculations of the vibrational and thermodynamic properties of NiX (X = S, Se, Te) compounds along with inelastic neutron scattering measurements of the phonon spectrum in NiSe. The measured phonon spectrum is in very good agreement with the computed result. We also report the measurement of thermal expansion behavior of NiSe using X-ray diffraction from 13 K to 300 K. The change in the hexagonal $c$ lattice parameter in NiSe is considerably greater as compared to $a$ parameter. The ab-initio calculated anisotropic Grüneisen parameters of the different phonon modes in all the chalcogenides along with the elastic constants are used to compute anisotropic thermal expansion behviour, which is found in good agreement with experiments. The displacement pattern of phonons indicate that difference in amplitudes of Ni and X atoms follow the anisotropy of thermal expansion behavior along c- and a-axis.

## I. INTRODUCTION

Transition metal chalcogenides, as important semiconductor materials, find technological applications in microelectronics[1-4]. Possible intercalation of these chalcogenides with lithium for applications in rechargeable battery electrodes has garnered a lot of attention. They are conventionally used in several high technological applications, like photovoltaics to name a few[4, 5,6]. The various electronic properties and characteristics are utilized in electronic and energy conversion devices. NiTe and NiS are commonly used in electronic industries for special resistors and capacitors[7]. These materials are used in electronic devices as gate and drain terminal and to reduce contact resistance at source[8]. Metal selenides are regarded as promising candidates for sodium ion anode materials owing to their good conductivity and high theoretical capacity[9]. Electrochemical water splitting has worldwide attention for its role in renewable alternate energy source. Splitting of water[10] in hydrogen and oxygen through application of sunlight or electric pulse has been an interesting avenue for this. NiSe finds its use as a catalyst in expediting this otherwise sluggish process[11]. They are a cheap and better alternate to the precious metal oxides like $IrO_2$ and $RuO_2$ which were traditionally used as electrocatalyst[12-14]. Nanocrystallites of NiTe, CoTe are used in electorchemical applications for determination of uric acid and adenine in biological fluids[15], which are otherwise very difficult to identify. NiS is also excellent in catalysis for hydrogenation[4], carbon liquefaction reactions, solar storage, electroactive material[16] for high performance supercapacitor. It is an attractive cathode material for rechargeable Li-batteries[16]. The usefulness of these chalcogenides has encouraged studies on their structural, electronic, thermal and transport properties[16-27]. Nano spheres of these compounds are studied and synthesized for several technological uses[28].

Monochalcogenides have incomplete d shell of the component transistion metal atoms which produces a variety of structural, magnetic, thermal and electrical properties. NiS is found in rhombohedral millerite phase (β-NiS, R3m)[29] at ambient conditions. β-NiS shows high metallic conducting Pauli paramagnet at room temperature[22]. In the millerite phase Ni atoms have 5 nearest S neighbours occupying corners of a square pyramid. At 652 K, millerite transforms to $P6_3/mmc$ (α-NiS, NiAs type structure))[29], which can be quenched to a metastable phase at room temperature. α-NiS undergoes a first order transition on cooling at 260 K, below which it is an antiferromagnet[17, 19]. This metal-insulator transition[23] has been of continued interest. Several spectroscopic studies on the metastable hexagonal NiS have been studied to understand this transition[20, 26]. The moments in hexagonal layers are coupled ferromagnetically and point along the c-axis. Adjacent hexagonal layers are antiferromagnetic. Average S-S distances in NiAs type NiS are shorter than those in millerite, hence the α-phase is denser. At room temperature volume per formula unit is smaller for α- NiS as compared to low temperature millerite.

NiSe and NiTe are found in the hexagonal phase at room temperature. There is ferromagnetic ordering below 20 K in NiSe and at about 130 K and 20 K in NiTe[1]. The electronic band structures in these systems are of interest to understand their catalytic activites. As the electronegativity of the chalcogen atom decreases down the series, degree of covalency in the metal-chalcogen bond increases (electronegativty of Te is less than Se)[10]. In this context nickel selenide exhibits best catalytic efficiencies. These compounds exhibit interesting magnetic and electrical properties. The crystal structure and magnetic properties of Ni Chalcogenides have been studied copiously[1, 22]. They form stable stoichiometric hexagonal structures, and also form other structures in both stoichiometric/non-stoichiometric forms. Accordingly they exhibit different magnetic ordering. FeSi, CoSi and similar compounds have been extensively studied with above applications in mind, and also for their various unusual physical and chemical properties. FeSi, a fascinating material, has been studied for its unusual magnetic and thermodynamic properties like thermal expansion, elastic behaviour, heat capacity, Seebeck coefficients, magnetic susceptibility and insulator-metal transition with increasing temperature[30]. Phonon density of states have been studied in FeSi, both[31] experimentally and using ab-initio molecular dynamics, to investigate softening of phonons and to understand the coupling between phonons and electronic structure in order to explain anomalous thermal expansion. Equation of state and elasticity of FeSi have been studied extensively to gain better understanding of the earth's core, which is mainly comprised of an alloy of iron and nickel, with a few light alloying elements like Si, C, S, H and O.

We have earlier conducted a detailed study of the phonon behavior and anomalous thermal expansion in NiSi and NiGe. Limited literature is available on selenides of Ni, Raman data on nanospheres[28] of NiSe have been reported, and no recent studies on the thermal expansion behavior of these compounds are reported. There have been several experimental and theoretical band structure calculations on NiS, while studies on NiTe is very scarce. Here, in this paper we report the thermal expansion of the NiSe from low temperature to room temperature by X-ray diffraction and also report the inelastic neutron scattering studies of the phonon density of states in NiSe. We have performed extensive ab-initio calculation of the lattice dynamics of NiX (X = S, Se, Te) in the hexagonal structure (space group P6$_3$/mmc). The results are used to understand and qualitatively derive anisotropic thermal expansion in these materials, in terms of specific anharmonic phonons.

## II. EXPERIMENTAL AND THEORETICAL DETAILS

Inelastic neutron scattering experiments to study the phonon density of states in polycrystalline NiSe sample have been carried out using the Triple Axis Spectrometer at Dhruva reactor, Trombay. A sample of about 20 grams of NiSe polycrystalline powder is used. All the measurements are carried out in the energy loss mode with constant momentum transfer (*Q*) geometry over a range of Q-values. The energy of the analyser in various scans was kept fixed at 25 meV or 34.4 meV. In the incoherent one-phonon approximation[32, 33], the measured scattering function *S(Q,E)*, as observed in the neutron experiments, is related to the phonon density of states $g^{(n)}(E)$ as follows:

$$g^{(n)}(E) = A < \frac{e^{2W(Q)}}{Q^2} \frac{E}{n(E,T) + \frac{1}{2} \pm \frac{1}{2}} S(Q,E) > \quad (1)$$

Where the + or − signs correspond to energy loss or gain of the neutrons respectively and $n(E,T) = \left[\exp(E/k_B T) - 1\right]^{-1}$.

The observed neutron-weighted phonon density of states is a sum of the partial components of the density of states due to the various atoms, weighted by their scattering length squares.

$$g^n(E) = B \sum_k \{\frac{4\pi b_k^2}{m_k}\} g_k(E) \quad (2)$$

Here $b_k$, $m_k$, and $g_k(E)$ are, respectively, the neutron scattering length, mass, and partial density of states of the $k^{th}$ atom in the unit cell. The quantity between <> represents suitable average over all *Q* values at a given energy. 2*W(Q)* is the Debye-Waller factor averaged over all the atoms. The weighting factors $\frac{4\pi b_k^2}{m_k}$ for the various atoms in the units of barns/amu are: Ni: 0.315, Te: 0.033 and Se: 0.105. *A* and *B* are normalization constants

X-ray diffraction studies were carried out using rotating Cu anode based powder diffractometer operating in the Bragg-Brentano focusing geometry with a curved crystal monochromator. Data were collected in the continuous scan mode at a scan speed of 2 degrees per minute at an interval of 0.02 degree in the 2θ range of 20°–130°. All the data collections were carried out during heating cycles of the sample for powder x-ray diffraction measurement in the temperature range 13 to 300 K. The accuracy of the measured temperature during data collection was within ±0.2 K. The structural refinements were performed using the Rietveld refinement program FULLPROF[34, 35].

The lattice dynamics calculations of NiX (X=S, Se and Te) have been performed using *ab-initio* methods. We have used the Vienna ab initio simulation package (VASP)[36, 37] along with the Phonopy[38] package for the *ab-initio* phonon calculations. The plane wave pseudo-potential has been used with maximum plane wave energy cutoff of 900 eV. The generalized gradient approximation (GGA)[39] exchange correlation given by Perdew, Becke and Ernzerhof with projected–augmented wave method has been used. The calculations are performed on a 20×20×20 k-point mesh in Brillouin zone generated by Monkhorst Pack method[40]. The valence electronic configurations of Ni and X (S, Se and Te), as used in calculations for pseudo-potential generation are $1s^2 2s^1$ and $s^2 p^4$ respectively.

The self consistent convergence criteria for the total energy and forces calculations were set to $10^{-8}$ eV and $10^{-4}$ eV Å$^{-1}$, respectively. The compounds NiX (X=S, Se and Te) are known to be magnetic order at low temperature. All the calculations are performed in the hexagonal phase (space group P6$_3$/mmc). The ferromagnetic ordering is considered for NiSe and NiTe, while A-type antiferromagnetic structre was used in the calculations for NiS. Further, NiS is known to be an insulator, so the on-site Hubbard correction is applied using the Dudarev approach[41]. We have used $U_{eff}$=4 eV as done in previous electronic structure calculations of NiS[19].

The phonon calculations are carried out in the relaxed configuration. That is the lattice constants as well as atomic coordinates of the atoms are fully optimized. The phonon calculation are perfomed using the supercell approach. A supercell of (3 × 3 × 2) dimension, which consist of 72 atoms has been used in the computations. The required force constants were computed by displacing atoms in different configurations of the supercell with (±x, ±y, ±z) atomic displacement patterns.

## III. RESULTS AND DISCUSSIONS

## A. Thermal Expansion Behaviour of NiSe Using X-ray Diffraction Technique

NiSe crystalizes in hexagonal structure with space group P6$_3$/mmc and lattice parameters are $a_h$= 3.6613 (Å), $c_h$= 5.3562 (Å), which is in very good agreement with literature[42]. It has two formula units per cell and has four atoms. Ni and Se atoms occupy 2a (0, 0,0) and 2c (1/3, 2/3, ¼) positions, respectively. To investigate thermal expansion behavior of NiSe at atomic level, we carried out temperature dependent powder X-ray diffraction. The diffraction data colleted at 13 K can indexed with hexagonal structure with space group P 6$_3$/mmc (see Figure 1 (a)) and suggest that NiSe does not undergoes any structural phase transition below room temperature. Structural parameters obtained by Rietveld analysis of temperature dependent powder diffraction data show that the lattice parameters monotonically increase with increasing temperature(Figure 1(b), and show anisotropic thermal expansion. The observed thermal expansion coefficient in the range of 10 - 300 K are $\alpha_a = 12.3 \times 10^{-6}$ K$^{-1}$ and $\alpha_c = 21.2 \times 10^{-6}$ K$^{-1}$

## B. Phonon Spectra

The comparison between the experimental and calculated structure of NiX (X=S, Se and Te) is given in Table I. Experimentally NiSe and NiTe are known to show very weak ferromagnetism at low temperatures with magnetic moment of ~0.004 µB[1]. The calculated magnetic moment for the equilibrium structure is NiSe and NiTe is 0.0 µB and 0.0 µB respectively. However NiS is know to undergo magnetic transition below 265 K[17] and have magnetic moment of 2.1 µB[17,22]. Our calculations give moment of 1.2 µB which is in agreement with previous ab-initio calculations.

Figure 2 gives the measured inelastic neutron scattering data of the phonon density of states in NiSe. It is compared to our ab-initio calculation, the agreement is excellent. The span of the phonon spectrum is only up to 30 meV. There are peaks around 8, 12, 18 and 26 meV. The spectrum is continuous. We have analyzed the measured neutron inelastic scattering data using the calculated phonon density of states.

Partial and total phonon density of states of constituent atoms of NiX compounds are given in figure 3. It is possible to gauge the contribution of the various atoms to the total phonon spectrum. In all the three compounds, Ni and X (S, Se, Te) contribute almost throughout the entire energy range. Total density of states in NiS extends upto 30 meV; in NiSe it is till 29 meV while in NiTe it extends till 27 meV. The energy span of Ni in NiS extends up to 30 meV with peaks at 13, 18 meV, 22 meV, 25 meV and 28 meV. In case of NiSe, the peaks in the Ni's spectra are found at 9, 12, 15, 18 , 22 and 27 meV. The span of the spectrum is up to 29 meV. In NiTe, Ni's peaks are at 11, 14, 17, 19, 21 and 25 meV.

The lowest energy peak in the spectra of X (S, Se, Te) atoms in S (32.06 amu) and Te (127.60 amu) compounds is at 13 meV, 8 meV and 10 meV respectively, which clearly shows that the peak position does not follow the mass renormalization indicating that the bonding of atoms is different among these compounds. The spectra of Se in NiSe extends up to 29 meV with peaks at 8, 16, 18, 19 and 24 meV respectively. In case of S in NiS, energy extends upto 30 meV with peaks at 13, 16, 21, 23, 26 and 28 meV. While in NiTe, peaks in the Te spectrum are at 10, 13, 17, 20 and 23 meV respectively.

The partial contribution of Ni and X (Se, Te) in NiSe and NiTe appears to be similar with very small differences. However, the lowest energy peak in the spectra of X (Se, Te) atoms in Se (78.96 amu) and Te (127.60 amu) compounds at 8 meV and 10 meV respectively does not follow mass normalization.

The calculated elastic constants of the compounds are given in Table II. $C_{11}$ of NiS is highest while that of NiTe is the lowest. $C_{33}$ of NiSe is higher than the other two chalcogenides. Bulk modulus of NiSe is highest, indicating its higher resistance to compression as compared to the other two chalcogenides. NiS is the most compressible of the three.

The group theoretical decomposition of phonons at the zone centre is given as:

$$\Gamma = 2A_{2u}+2E_{1u}+E_{2u}+E_{2g}+B_{2g}+B_{1u}$$

The computed zone centre phonons have been given in Table III[26]. The agreement with the experimentally reported values seems fair.

## C. Thermal Expansion Behavior

The anisotropic thermal expansion behaviour can be computed under the quasiharmonic approximation. Each phonon mode of energy $E_{qj}$ ($j^{th}$ phonon mode at point q in the Brillouin zone) contributes to the thermal expansion coefficient. The anisotropic linear thermal expansion[43] coefficients are given by:

$$\alpha_l(T) = \frac{1}{V_0} \sum_{q,i} C_V(q,i,T) [s_{l1}\Gamma_a + s_{l2}\Gamma_b + s_{l3}\Gamma_c], \quad l = a,b,c$$

Where $V_0$ is volume at 0 K, $s_{ij}$ are elements of elastic compliances matrix (which is inverse of the elastic constant matrix) at constant temperature (0 K) and $C_V(q, i, T)$ is the specific heat at constant volume for $i^{th}$ phonon mode with wave vector **q** in the Brillouin zone.

The calculation of the anisotropic thermal expansion behavior in a compound requires the information of the anisotropic linear mode Grüneisen parameters ($\Gamma_a, \Gamma_b$ and $\Gamma_c$). The effect of change in volume of a compound on its vibrational properties is related by mode Grüneisen parameter ($\Gamma$). We have obtained the Grüneisen parameters from pressure dependence of phonon frequencies in entire Brillouin zone. $\Gamma_a, \Gamma_b$ and $\Gamma_c$ are the anisotropic linear mode Grüneisen parameters of phonon of energy $E_{q,j}$ is given by,

$$\Gamma_l(E_{q,i}) = \left(-\frac{\partial \ln E_{q,i}}{\partial \ln l}\right)_{T,l'}; \quad l, l' = a, b, c \;\&\; l \neq l' \quad (1)$$

For hexagonal system $\Gamma_a = \Gamma_b$. The volume thermal expansion coefficient for a hexagonal system is given by:

$$\alpha_V = (2\alpha_a + \alpha_c) \quad (2)$$

The calculations of $\Gamma_l(E)$ are performed by applying an anisotropic stress by changing the lattice constant 'a' and keeping the 'b' and 'c' parameters constant; and vice versa following eq.(1). The calculated variation of anisotropic Grüneisen parameters with energy averaged over all the phonon modes in the Brillouin zone on application of anisotropic stress along a and c directions in NiS, NiSe and NiTe is shown in Figure 4.

In NiS, $\Gamma_a$ is less than $\Gamma_c$ till 21 meV, between 21 to 23 meV, $\Gamma_a$ is slightly more than $\Gamma_c$; and beyond that $\Gamma_a$ reduces while $\Gamma_c$ increases. In case of NiSe, $\Gamma_a$ is less than $\Gamma_c$ till 10 meV; $\Gamma_a$ is greater between 10 and 15 meV. Between 16 meV to 22 meV, both are comparable, while beyond 22 meV, $\Gamma_c$ increases two fold. In NiTe, $\Gamma_a < \Gamma_c$ up to 10 meV; $\Gamma_c < \Gamma_a$ between 10-15 meV, they are almost same between 16-21 meV; $\Gamma_c$ increase two fold from 22 meV. The average value of $\Gamma_a$ and $\Gamma_c$ in NiS is lower in comparison to that in NiSe and NiTe.

The calculated thermal expansion coefficients are shown in Fig. 5. We find that for all the three chalcogenides, NiS, NiSe and NiTe, the linear thermal expansion coefficients $\alpha_l$ ($l$=a,c) along a- and c- axis are found to be positive in the whole temperature range. In NiS, $\alpha_c$ is almost four times $\alpha_a$. In case of NiSe, $\alpha_c$ is approximately twice $\alpha_a$. In NiTe both $\alpha_c$ and $\alpha_a$ are comparable.

The comparison between our experimental and calculated change in the lattice parameters with increasing temperature in case of NiSe is given in Figure 6, along with change in unit cell volume. In case of NiS and NiTe, the calculated changes are in good agreement with reported experimental data[25,29]. The agreement between our experiment and calculation in case of NiSe is excellent. The relative change in the c- parameter is greater as compared to that in the a- parameter.

The thermal expansion coefficient at a given temperature has contribution of phonon modes of different energies in the entire Brillouin zone. The contribution to anisotropic linear thermal expansion at 300 K from a phonon of energy E in the entire Brillouin zone is shown in Figure 7. In NiSe and NiTe, the mode specific thermal expansion along a- axis is positive from modes up to 26 meV. The high energy modes around 28 meV contribute to to small negative to $\alpha_a$. While in NiS, it is slightly negative from phonons in a very small range between 10 and 15 meV. As far the behavior along c- axis is concerned, in case of NiS the contribution is positive throughout. However, both in NiSe and NiTe there is a small region between 10 and 15 meV, wherein the contribution to thermal expansion along c-axis is negative. It is interesting to note that the overall volume thermal expansion is always positive in all the three compounds.

As shown in Fig. 7, the maximum anisotropy in $\alpha_c/\alpha_a$ is for NiS. In NiS, in sync with the trend displayed by $\Gamma(E)$; the contribution to $\alpha_c$ from phonons in the entire energy range up to 30 meV is greater than that in $\alpha_a$. The phonons around 12 meV contribute a very small negative value to $\alpha_a$. The anisotropy in $\alpha_c/\alpha_a$ behavior in NiSe and NiTe seems to be reduced in comparison to that in NiS. As shown in Fig. 7 in both NiSe and NiTe, the low energy phonons upto 10 meV contribute much more to $\alpha_c$ in comparison to that to $\alpha_a$. Further it can be seen that in NiSe the contribution to $\alpha_c$ from 8 meV phonons in more than that in NiTe. Basically the difference in contribution to $\alpha_c$ from modes around 8 meV results in difference in anisotropic behavior of $\alpha_c/\alpha_a$ in NiSe and NiTe. Further it can be noticed that the large difference in anisotropy in $\alpha_c/\alpha_a$ in NiS and other two compounds (NiSe and NiTe) is due to difference in contribution to $\alpha_a$ from modes around 12 meV. The large difference in contribution from these modes results in anisotropy in $\alpha_c/\alpha_a$ in these compounds.

In order to understand the nature of phonons responsible for large anistotropic thermal expansion behavior we have calculated the contribution of phonons of energy E to ansitropic mean square displacement of Ni and X (X = S, Se, Te) averaged over Brillouin zone at 300 K (Fig 8). Due to the hexagonal symmetry x and y components are the same. The structure of NiX consists of hexagonal sheets formed by Ni and X atoms.

In case of NiS, the contribution to $u^2$ of S atoms, along a- axis, from all the phonons in the entire energy range is greater than that of $u^2$ of Ni. However, along c- axis, the contributon of S atoms in the range of 10-15 meV is substantially higher than that of Ni. As shown in Fig.7, the phonons of energy around 12 meV, which have large contribution from S atoms, contribute maximum to $\alpha_c$.

For NiSe, $u_x^2$ of Ni is throughout higher and greater than that of Se. However, for phonons of energy around 10 meV, Ni atoms have significantly large amplitude in comparison to Se. Similarly $u_z^2$ of various atoms indicate that phonons of energy about 8 meV, 15 meV and 26 meV involve large amplitudes of Se, Ni and Ni respectively. The phonons involving large amplitudes along a and c directions seem to make large contribution to $\alpha_a$ and $\alpha_c$ respectively.

In NiTe, mean square amplitude of Ni along a- axis is much greater than that of Te up to 18 meV. The Ni amplitude along a-axis peaks at about 10 meV. Along c- axis, $u^2$ of Ni is more than that of Te and peaks at 10 meV, 16 meV and 26 meV, which also correspond to peaks in $\alpha_c$.

Further, as shown in Fig. 5, the maximum value of $\alpha_c$ at 300 K among the three compounds is for NiS, while NiTe has least $\alpha_c$. As shown in Fig. 7, the ratio of amplitude of $u_z^2$ of X/Ni or Ni/X for phonons of 12 meV is maximum for NiS and least in NiTe. Similarly $\alpha_a$ values for NiSe and NiTe are nearly same at 300 K, while for NiS the $\alpha_a$ value is least. The ratio of amplitude of $u_x^2$ of X/Ni or Ni/X for phonons of 10 meV follows similar trend, indicating nearly same and large value for NiSe and NiTe and least for NiS.

The anistropic mean squared displacement of various atoms (Fig 8) gives the average nature of dynamics. In order to understand the microscopic mechanism of anistropic thermal expansion behaviour we have plotted displacement patterns of zone centre modes (Fig. 9).

The lowest energy $E_{1u}$ zone centre optic mode involves (Fig. 9) the displacement of Ni atoms along the a-axis, while X atoms are at rest. This mode has large anisotropy in the Grueneisen parameters that gives rise to negative and positive thermal expansion along a-axis and c-axis respectively. The second zone centre $E_{2g}$ optic mode (Fig. 9) has positive contribution to $\alpha_a$ and $\alpha_c$. In this mode Ni atoms are at rest while X atoms connected to vertices of hexagon move in opposite directions. The $A_{2u}$ mode involves displacement (Fig. 9) of Ni atoms along c-axis, while X atoms have displacement opposite to the movement of Ni atoms.

The fourth zone centre $B_{2g}$ optic mode involves antiparallel displacements (Fig. 9) of nearest Ni atoms along c-axis and contribute positive thermal expansion to $\alpha_a$ and $\alpha_c$. The fifth mode of $E_{1u}$ symmetry contributes large anisotropic components to $\alpha_a$ and $\alpha_c$. In this mode Ni atoms show displacements (Fig. 9) along the b-axis, while X-atoms move in a direction opposite to Ni along the b-axis. The highest energy $B_{1u}$ mode involes (Fig. 9) displacement of Ni atoms towards c-axis while X atoms are at rest. The mode has large positive contribution to $\alpha_c$ and small negative contribution to $\alpha_a$.

## IV. CONCLUSIONS

In this paper, we report a comprehensive study of X-ray diffraction and inelastic neutron scattering experiments, lattice dynamics calculation in NiSe; along with theoretical lattice dynamics simulations in NiS and NiTe. The interest in these compounds originated from its various technological uses. Our ab-initio lattice dynamics studies in NiSe are found to be in very good agreement with the measured data of phonon density of states. We find slight differences in the partial densities of Ni and X (X = S, Se and Te) for all the three compounds, although they are structurally similar and have similar spectra. Our experimental measurement of the thermal expansion in NiSe is in very good agreement with the computed data. The difference in anistopy in thermal expansion coefficients in the three compounds is explained in terms of atomic displacement patterns as well as ratio of amplitudes of X and Ni atoms.

TABLE I. Comparison between the experimental[1, 29] (at 293 K) and calculated structural parameters (at 0 K) of NiX (X = S, Se,Te). For hexagonal structure of NiX (space group P6$_3$/mmc) the Ni and the S atoms occupy Wyckoff positions (0,0,0) and (1/3, 2/3, 1/4) respectively and their symmetry equivalent positions

|  | NiS | | NiSe (our work) | | NiTe | |
| --- | --- | --- | --- | --- | --- | --- |
|  | Experimental | Calculated | Experimental | Calculated | Experimental | Calculated |
| $a$ (Å) | 3.4395 | 3.4781 | 3.6613 | 3.7028 | 3.96 | 4.0279 |
| $c$ (Å) | 5.3514 | 5.3889 | 5.3562 | 5.2201 | 5.35 | 5.2888 |

Table II: Calculated elastic constants along with bulk modulus (B) in GPa units.

| Elastic constant | NiS | NiSe | NiTe |
| --- | --- | --- | --- |
| $C_{11}$ | 168 | 162 | 157 |
| $C_{12}$ | 70 | 111 | 86 |
| $C_{13}$ | 63 | 85 | 69 |
| $C_{33}$ | 153 | 172 | 149 |
| $C_{44}$ | 38 | 19 | 28 |
| B | 98 | 117 | 101 |

Table III: The calculated Raman and infra red modes of NiX (X = S, Se,Te) in cm$^{-1}$ units

|  | NiS | NiSe | NiTe |
| --- | --- | --- | --- |
| $E_{2u}$ | 125 | 93 | 92 |
| $E_{2g}$ | 159 | 141 | 139 |
| $A_{2u}$ | 173 | 144 | 140 |
| $B_{2g}$ | 181 | 173 | 190 |
| $E_{1u}$ | 205 | 163 | 165 |
| $B_{1u}$ | 229 | 212 | 194 |

**Fig 1.** (Color online) The results of Rietveld refinement fitting of the observed (circles) and calculated (solid line) powder X-ray diffraction data using hexagonal crystal structure with space group ***P6₃/mcm*** at 13 K. Figure 1 (b) shows the variation of lattice parameters with temperature.

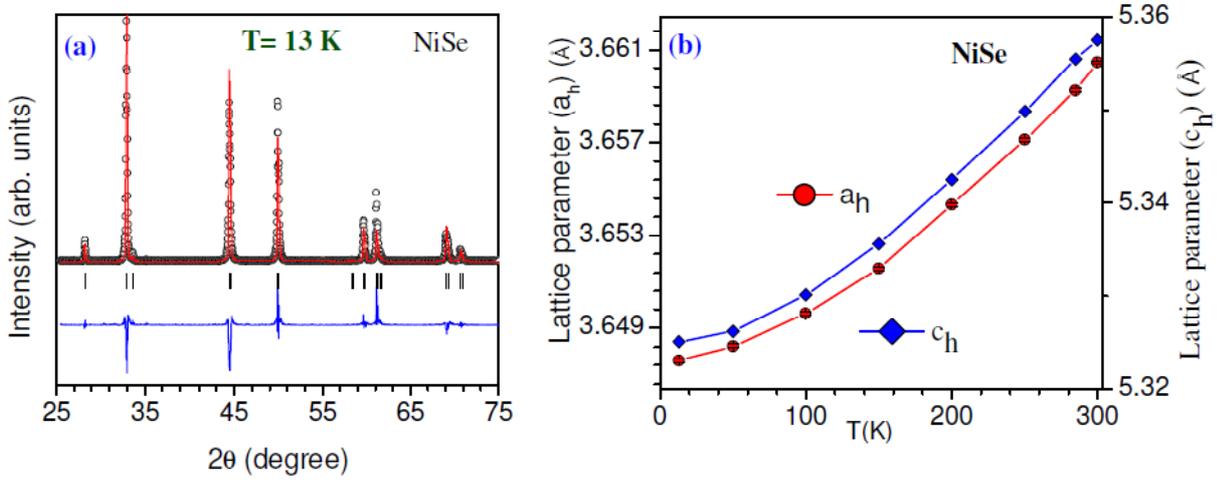

FIG 2 (Color online): Inelastic neutron scattering phonon spectrum (line with symbols) compared with our computed neutron-weighted phonon density of states at ambient conditions.

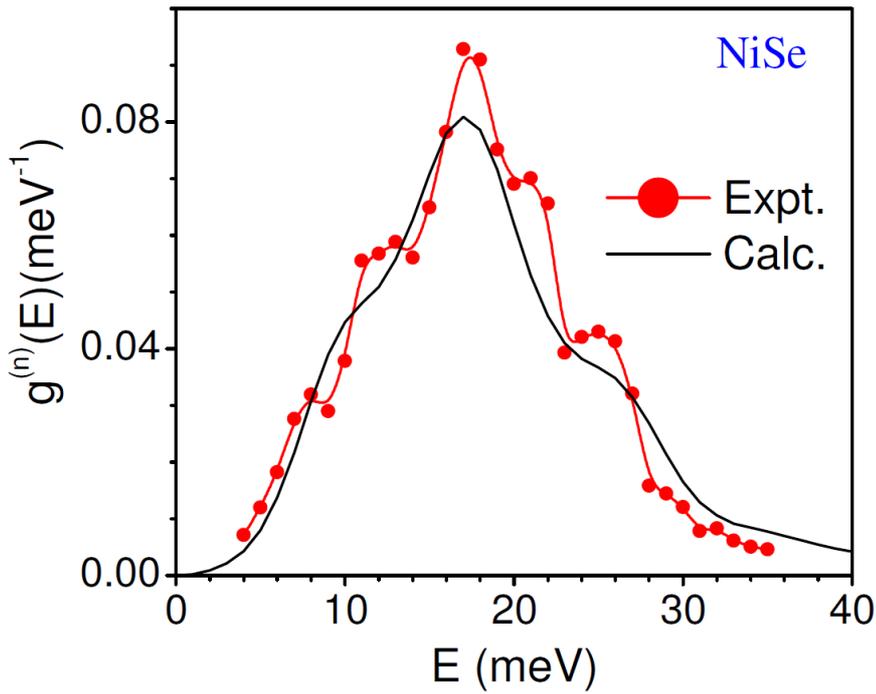

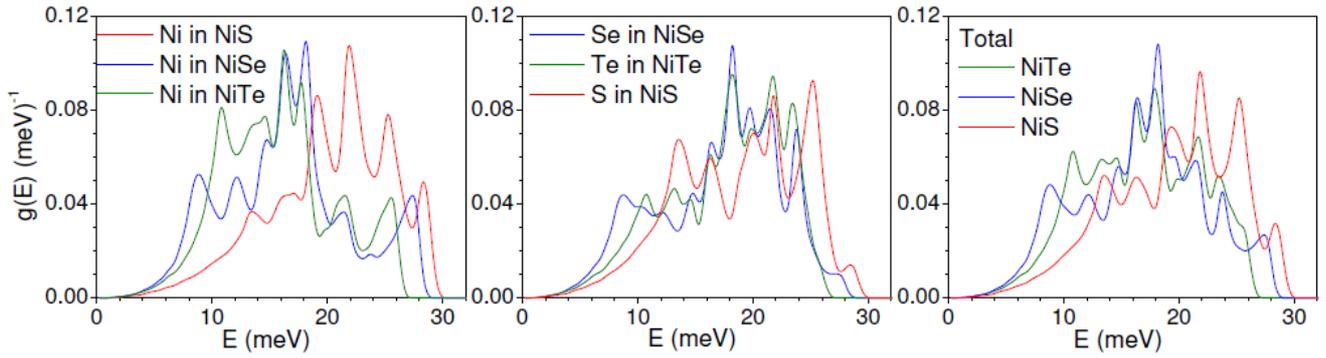

FIG 3 (Color online): Computed partial and total phonon density of states in NiS, NiSe and NiTe.

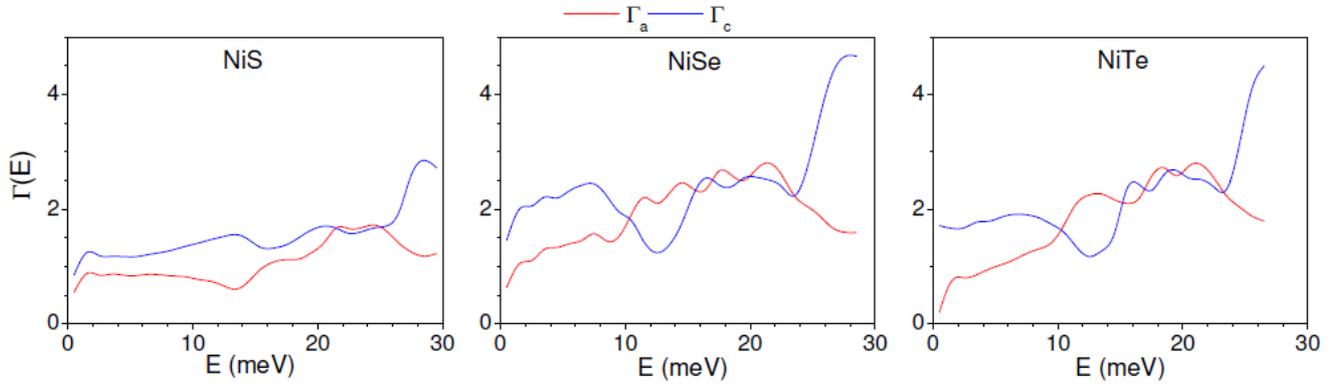

FIG 4 (Color online): Variation of the anisotropic Grüneisen parameters with energy in the three chalcogenides.

FIG 5 (Color online): Calculated anisotropic thermal expansion in the chalcogenides.

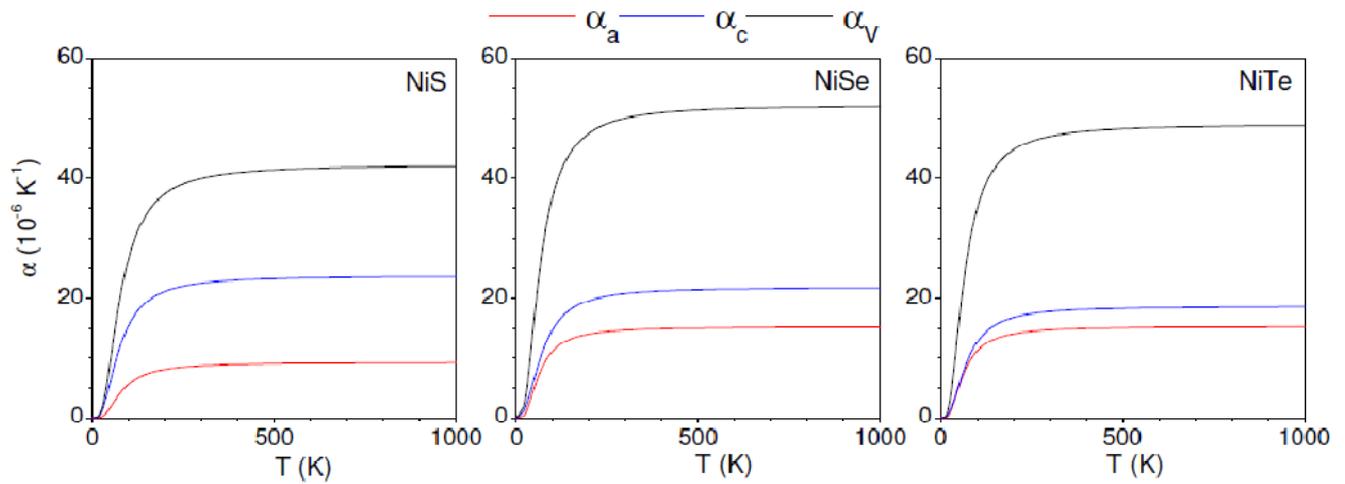

FIG 6 (Color online): Variation of *a* and *c* parameters, volume (X= a, c, V) with temperature in NiSe, NiS and NiTe at ambient pressure. The experimental studies on NiSe has been carried out by us, while data on NiS and NiTe is from literature[25, 29].

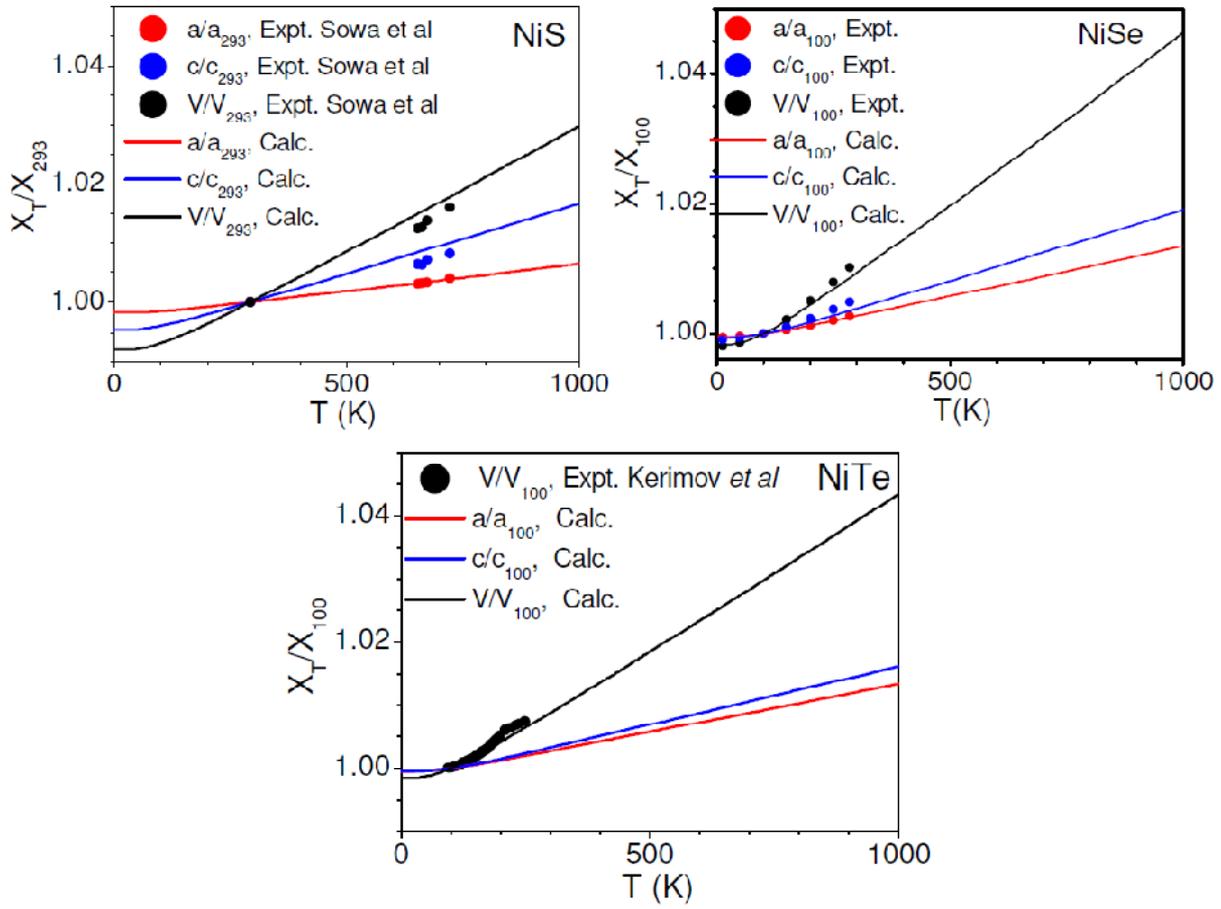

FIG 7 (Color online): The contribution to anisotropic linear thermal expansion coefficients at 300 K, from phonon mode of energy E averaged over the Brillouin zone.

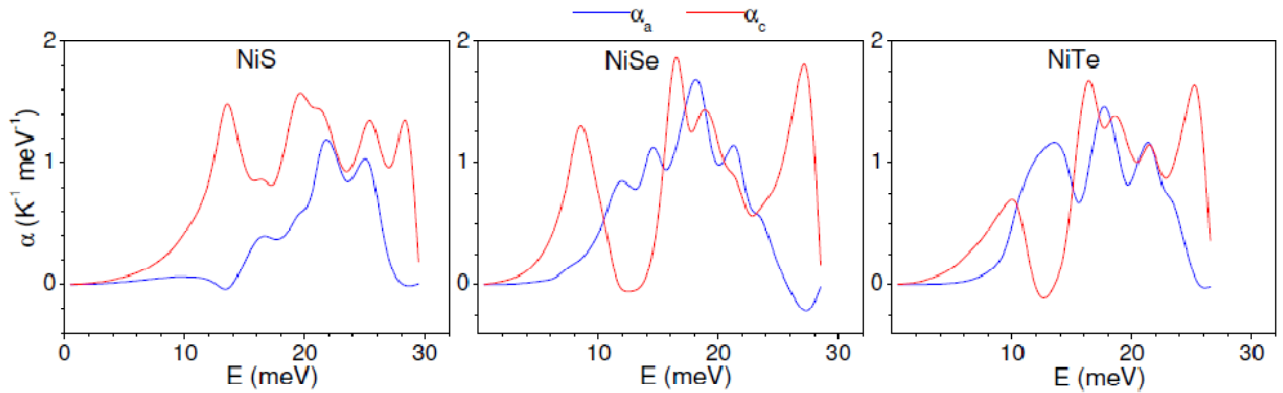

FIG 8 (Color online): The calculated anisotropic mean square displacement of Ni and X (X = S, Se, Te) as a function of phonon energy E averaged over Brillouin zone at 300 K.

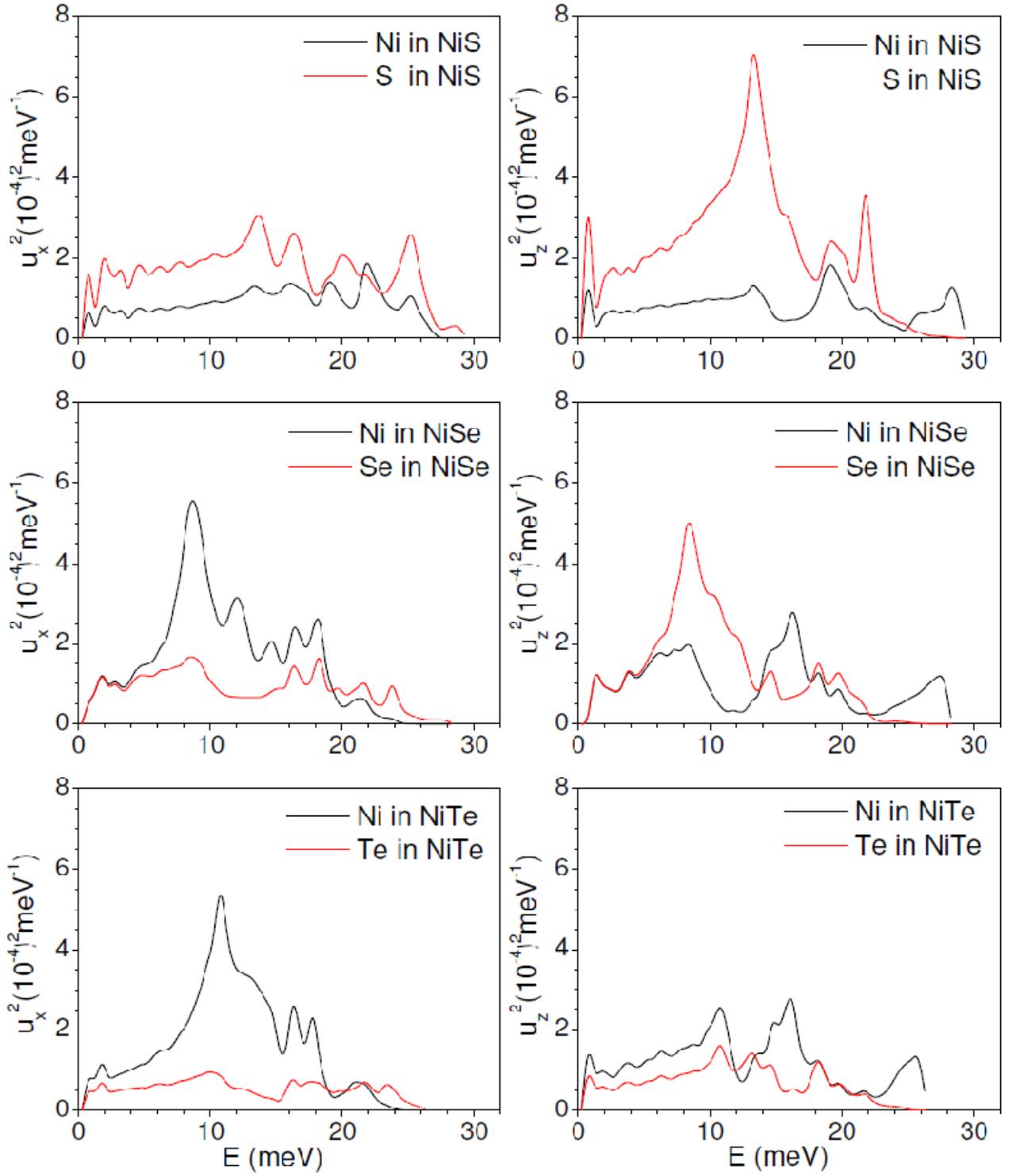

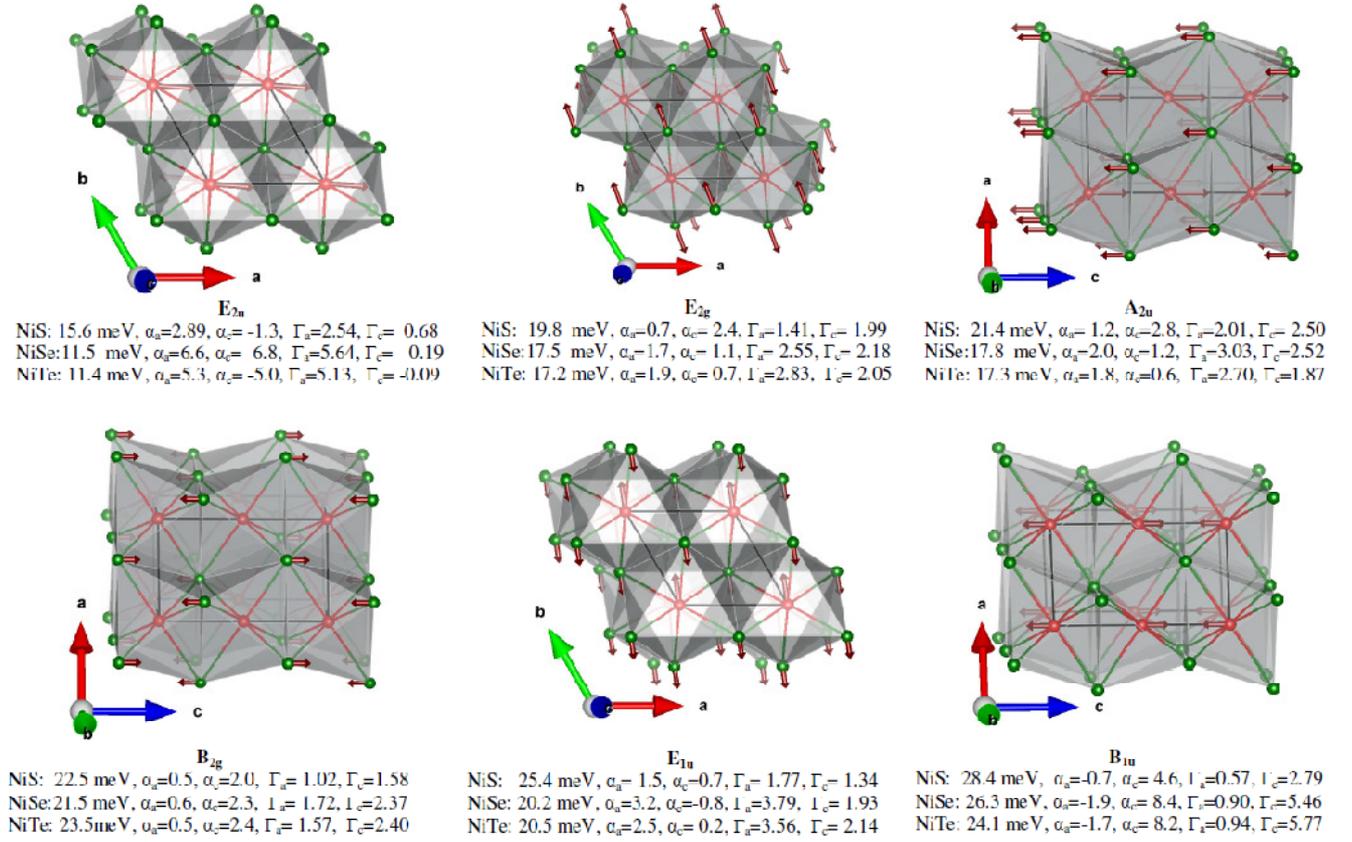

FIG 9 (Color online): The atomic displacement pattern of zone centre Phonon modes in the NiX (X = S, Se, Te). The values of linear thermal expansion coefficients ($\alpha_a$ and $\alpha_c$) are at 300 K and are in the units of $10^{-6}$ K$^{-1}$. Key- Ni: red, X:green.